\documentclass[twocolumn]{el-author}

\usepackage{epstopdf}
\usepackage[normalem]{ulem}

\newcommand{\ua}{\uparrow}
\newcommand{\nc}{\newcommand}
\makeatletter
\newcommand{\vast}{\bBigg@{1.75}}
\newcommand{\Vast}{\bBigg@{2.4}}
\newcommand{\vastl}{\bBigg@{4}}
\newcommand{\Vastl}{\bBigg@{5}}
\makeatother
\nc{\da}{\downarrow} \nc{\hc}{\hat{c}} \nc{\hS}{\hat{S}}
\nc{\bra}{\langle} \nc{\ket}{\rangle} \nc{\eq}{equation (\ref}
\nc{\h}{\hat} \nc{\hT}{\h{T}}\nc{\be}{\begin{eqnarray}}
\nc{\ee}{\end{eqnarray}}\nc{\rd}{\textrm{d}}\nc{\e}{eqnarray}\nc{\hR}{\hat{R}}\nc{\Tr}{\mathrm{Tr}}
\nc{\tS}{\tilde{S}}\nc{\tr}{\mathrm{tr}}\nc{\8}{\infty}\nc{\lgs}{\bra\ua,\phi|}\nc{\rgs}{|\ua,\phi\ket}
\nc{\hU}{\hat{U}}\nc{\lfs}{\bra\phi|}\nc{\rfs}{|\phi\ket}\nc{\hZ}{\hat{Z}}\nc{\hd}{\hat{d}}\nc{\mD}{\mathcal{D}}
\nc{\bd}{\bar{d}}\nc{\bc}{\bar{c}}\nc{\mc}{\mathcal}\nc{\ea}{eqnarray}\nc{\mG}{\mathcal{G}}\nc{\bce}{\begin{center}}
\nc{\ece}{\end{center}}
\date{21th March 2018}

\begin{document}

\title{Unified approaches based effective capacity analysis over composite $\alpha-\eta-\mu$/gamma fading channels}

\author{H. Al-Hmood and H. S. Al-Raweshidy}

\abstract{This letter analyses the effective capacity of communications system using unified models. In order to obtain a simple closed-form mathematically tractable expression, two different unified approximate models have been used. The mixture gamma (MG) distribution which is highly accurate approximation approach has been firstly employed to represent the signal-to-noise-ratio (SNR) of fading channel. In the second approach, the mixture of Gaussian (MoG) distribution which is another unified representation approach has been utilised. A comparison between the simulated and numerical results using both distributions over composite $\alpha-\eta-\mu$/gamma fading channels has been provided.}

\maketitle

\section{Introduction}
The ergodic capacity that is proposed by Shannon is measured by assuming no delays for wireless communication systems. Therefore, the effective capacity (or effective rate) has been suggested as a performance metric that can be used to measure the system behaviour under the quality of service restrictions such as system delays [1]. In the effective capacity, guaranteed statistical delay restrictions are assumed to be presented when the maximum constant value of the throughput that arrives at the transmitter is measured. Accordingly, several studies have been devoted to analyse the effective capacity over wireless fading channels [2]. To represent the line-of-sight (LoS), non-LoS (NLoS), and non-linearity communication scenarios of wireless fading channels, the $\kappa-\mu$, $\eta-\mu$, and $\alpha-\mu$ distributions which are generalised models that provide better practical results than the traditional distributions such as Nakagami-$m$ are investigated in [3], [4], and [5], respectively.
\par The impact of shadowing fading is also considered in the analysis of the effective capacity of communication systems over composite fading channels such as generalised-$K$ and Weibull/gamma [2, 6]. In [7], the $\kappa-\mu$ shadowed fading channel which is composite of $\kappa-\mu$ and Nakagami-$m$ distributions is utilised to model the fading channel. However, no works have been dedicated to analyse the effective capacity over composite $\eta-\mu$/shadowing and $\alpha-\mu$/shadowing fading channels. Furthermore, the unified framework in [6] is based on the moment generating function (MGF) of the instantaneous signal-to-noise ratio (SNR) that cannot be obtained in exact closed-form expression.
\par Motivated by there is no general unified approach for the effective capacity, this letter provides two different frameworks by using mixture gamma (MG) [8] and mixture of Gaussian (MoG) distributions [9]. These distributions have been widely utilised in the analysis of digital communication systems [10, 11]. This is because they provide simple closed-form analytic expression of the performance metrics. To this effect, the effective capacity over composite $\alpha-\eta-\mu$/gamma fading condition which is more generalised than the aforementioned channels is analysed using MG and MoG distributions. The main difference between the MG and MoG distributions is the number of the parameters that is required to achieve a minimum mean square (MSE) between the probability density function (PDF) of the exact and approximate models.
   
\section{System model}
The normalised effective capacity over fading channels is expressed by [7, eq. (1)]
\label{eqn_1}
\begin{equation}
\mathcal{R}=-\frac{1}{A}\mathrm{log}_2\big(\mathbb{E}\{(1+\gamma)^{-A}\}\big)
\end{equation}
where $\mathbb{E}\{.\}$ stands for the expectation and $A \triangleq \theta TB/\mathrm{ln}2$ with $\theta$, $T$, and $B$ denote the delay exponent, block duration, and bandwidth of the system, respectively. 
\section{The MG distribution}
Using a MG distribution, the PDF of the instantaneous SNR can be written as [8, eq.(1)]
\begin{equation}
f_\gamma(\gamma)=\sum_{l=1}^S \phi_l \gamma^{\vartheta_l-1} e^{-\xi_l \gamma}
\end{equation}
where $S$ is the number of Gamma distributions which is obtained via calculating the minimum MSE between (2) and exact PDF and $\phi_l$, $\vartheta_l$, and $\xi_l$ correspond to the parameters of $l$ Gamma component. 
\section{The MoG distribution}
The PDF of the instantaneous SNR, $\gamma$, can be expressed using a MoG as [9, eq. (24)]
\label{eqn_3}
\begin{equation}
f_\gamma(\gamma)=\sum_{i=1}^N \frac{\rho_i}{\sqrt{8\pi \bar{\gamma}\gamma}\psi_i}e^{-\frac{\big(\sqrt{\frac{\gamma}{\bar{\gamma}}}-\upsilon_i\big)^2}{2\psi^2_i}}
\end{equation}
where $N$ is the number of Gaussian components that provides minimum MSE and $\rho_i$, $\psi_i$, and $\psi^2_i$ are the weight, mean, and variance of the $i$th component, respectively. Moreover, $\sum_{i=1}^N \rho_i =1$ with $\rho_i > 0$.
\section{MG distribution based analysis}
It can be noted that (1) can be expressed as
\label{eqn_4}
\begin{equation}
\mathcal{R}=-\frac{1}{A}\mathrm{log}_2\bigg(\int_0^\infty(1+\gamma)^{-A}f_\gamma(\gamma)d{\gamma}\bigg)
\end{equation}
\par Substituting (2) in (4), this yields
\label{eqn_5}
\begin{equation}
\mathcal{R}=-\frac{1}{A}\mathrm{log}_2 \bigg(\sum_{l=1}^S \phi_l \int_0^\infty(1+\gamma)^{-A}\gamma^{\vartheta_l-1} e^{-\xi_l \gamma}d{\gamma}\bigg)
\end{equation}
\par Employing [2, eq. (9)] to compute the integration in (4), the following unified closed-from is obtained
\label{eqn_6}
\begin{equation}
\mathcal{R}=-\frac{1}{A}\mathrm{log}_2 \bigg(\sum_{l=1}^S \phi_l \Gamma(\vartheta_l) U(\vartheta_l;\vartheta_l+1-A;\xi_l)\bigg)
\end{equation}
where $\Gamma(.)$ is the incomplete Gamma function and $U(.;.;.)$ is the Tricomi hypergeometric function defined in [12, eq. (39)].
\section{MoG distribution based analysis}
When (3) is inserted in (4), the integral cannot be solved in exact closed-form. Accordingly, we express the functions of the integral in terms of Meijer G-function by using [13, eq. (10)], [13, eq. (11)], and [14, eq. (01.03.26.0115.01)] 
\label{eqn_7}
\begin{align}
(1+\gamma)^{-A}&=\frac{1}{\Gamma(A)}
G^{1,1}_{1,1} \Vast[\begin{matrix}
    1-A\\\\
  0\\ 
\end{matrix}\Vast\vert\gamma
\Vast]\nonumber\hspace{-0.8 cm}\\
e^{-\frac{\gamma}{2\bar{\gamma}\psi^2_i}}&=G^{1,0}_{0,1} \Vast[\begin{matrix}
  -\\\\
  0\\
\end{matrix}\Vast\vert\frac{\gamma}{2\psi^2_i \bar{\gamma}}
\Vast] \nonumber\hspace{-0.8 cm}\\
e^{\frac{\sqrt{\gamma}\upsilon_i}{\sqrt{\bar{\gamma}}\psi^2_i}}&=2 \pi^\frac{3}{2}G^{2,0}_{2,4} \Vast[\begin{matrix}
    0.25,0.75\\\\
  0,0.5,0.25,0.75\\ 
\end{matrix}\Vast\vert\frac{\upsilon^2_i\gamma}{4\psi^4_i \bar{\gamma}}
\Vast]
\end{align}
where $G^{a,b}_{n,m}[.]$ is the Meijer's G-function [15, eq. (7)]. 
\par Plugging (7) in (4), this yields 
\label{eqn_8}
\begin{align}
\mathcal{R}&=-\frac{1}{A}\mathrm{log}_2 \Vast(\sum_{i=1}^M \frac{\rho_i \pi}{\sqrt{2 \bar{\gamma}}\psi_i \Gamma(A)}e^{-\frac{\upsilon_i^2}{2\psi^2_i}}\int_0^\infty\gamma^{-\frac{1}{2}}G^{1,0}_{0,1} \Vast[\begin{matrix}
  -\\\\
  0\\
\end{matrix}\Vast\vert\frac{\gamma}{2\psi^2_i \bar{\gamma}}
\Vast] \nonumber\\ 
&\times G^{2,0}_{2,4} \Vast[\begin{matrix}
    0.25,0.75\\\\
  0,0.5,0.25,0.75\\ 
\end{matrix}\Vast\vert\frac{\upsilon^2_i\gamma}{4\psi^4_i \bar{\gamma}}
\Vast]G^{1,1}_{1,1} \Vast[\begin{matrix}
    1-A\\\\
  0\\ 
\end{matrix}\Vast\vert\gamma
\Vast]
d\gamma\Vast)
\end{align}
\par With the aid of [15, eq. (9)], the integral in (8) can be computed in exact closed-form as follows 
\label{eqn_9}
\begin{align}
\mathcal{R}&=-\frac{1}{A}\mathrm{log}_2 \Vast(\sum_{i=1}^M \frac{\rho_i \pi}{\sqrt{2 \bar{\gamma}}\psi_i \Gamma(A)}e^{-\frac{\upsilon_i^2}{2\psi^2_i}}
\hspace{0.5 cm}\nonumber\hspace{-0.9 cm}\\ 
&\times G^{1,1:1,0:2,0}_{1,1:0,1:2,4} \Vast[\begin{matrix}
    0.5\\\\
  A-0.5\\ 
\end{matrix}\Vast\vert
\begin{matrix}
    -\\\\
  0\\ 
\end{matrix}\Vast\vert
\begin{matrix}
    0.25,0.75\\\\
  0,0.5,0.25,0.75\\ 
\end{matrix}\Vast\vert
\frac{1}{2 \psi^2_i \bar{\gamma}}, \frac{\upsilon_i^2}{4 \psi^4_i \bar{\gamma}}
\Vast]\Vast)
\end{align}
\par It can be observed that (9) includes a Meijer's $G$-function of two variables which can be evaluated by using the MATHEMATICA program that is implemented in [15]. 
\section{The PDF of $\alpha-\eta-\mu$/gamma fading channels}
The PDF of SNR in $\alpha-\eta-\mu$/gamma fading can be calculated via averaging the PDF of $\alpha-\eta-\mu$ [16,  eq. (8)] over Gamma distribution as follows    
\label{eqn_10}
\begin{align}
f_\gamma(\gamma)&=\frac{\sqrt{\pi} \alpha h^\mu \mu^{\mu+\frac{1}{2}} \gamma^{\frac{\alpha}{2}(\mu+\frac{1}{2})-1}}{\Gamma(\mu) \Gamma(b) \Omega^b H^{\mu-\frac{1}{2}}} \nonumber\\
&\times \int_0^\infty x^{b-\frac{\alpha}{2}(\mu+\frac{1}{2})-1}e^{-\frac{2 \mu h \gamma^{\frac{\alpha}{2}}}{x^{\frac{\alpha}{2}}}-\frac{x}{\Omega}} I_{\mu-\frac{1}{2}}\bigg(\frac{2 \mu H \gamma^{\frac{\alpha}{2}}}{x^{\frac{\alpha}{2}}}\bigg)dx
\end{align} 
where $\alpha$, $\mu$, $b$, and $\Omega$ stand for the non-linearity severe parameter, the number of multipath clusters, shadowing index, and  mean power, respectively. Moreover, where $I_{k}(.)$ is the modified Bessel function of the first kind and $k$th order [12]. The parameters $H$ and $h$ are related to $\eta$ which represents a relationship between the quadrature and in-phase scattered components, into two formats, format 1 and format 2. In the former, $H=(\eta^{-1}-\eta)/4$ and $h=(2+\eta^{-1}+\eta)/4$ with $\eta\in(0, \infty)$ denotes the power ratio between the components whereas in the latter $H=\eta/(1-\eta^2)$ and $h=1/(1-\eta^2)$ with $\eta\in(-1, 1)$ refers to the correlation coefficient between the components [16].
\par Using $z=\frac{2 \mu h \gamma^{\frac{\alpha}{2}}}{x^{\frac{\alpha}{2}}}$ and following the same steps in [11], (10) can be expressed by a MG distribution with the following parameters 
\label{eqn_11}
\begin{align}
\phi_l=\frac{\theta_l}{{\sum_{j=1}^{S}}\theta_j \Gamma(\vartheta_j) {\xi_j}^{-\vartheta_j}}, \quad
 \vartheta_l=b, \quad \xi_l=\frac{(2 \mu h)^{\frac{\alpha}{2}}}{\Omega z^{\frac{\alpha}{2}}_l},\nonumber\\
 \theta_l=\frac{\sqrt{\pi} 2^{\frac{2}{\alpha}b-\mu+\frac{1}{2}} \mu^{\frac{2}{\alpha}b} h^{\frac{2}{\alpha}b-\frac{1}{2}}}{\Gamma(\mu) \Gamma(b) \Omega^b H^{\mu-\frac{1}{2}}} w_l z^{\mu-\frac{2}{\alpha}b-\frac{1}{2}}_l I_{\mu-\frac{1}{2}}\vast(\frac{H}{h} z_l\vast)
\end{align}

\begin{figure}[h]
\centering
  \includegraphics[width=3.5 in, height=2.32 in]{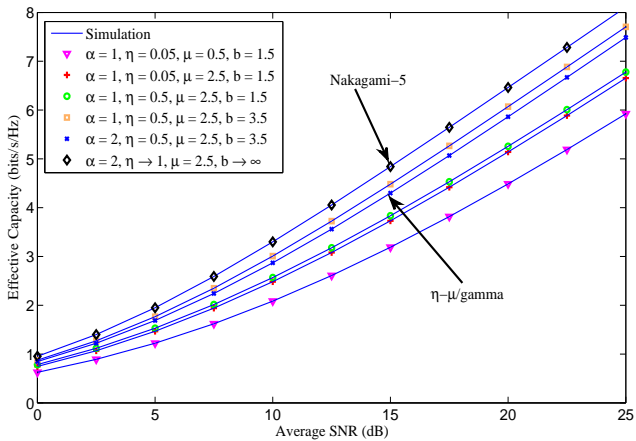} 
\centering
\caption{Simulated and numerical effective capacity using MG distribution against average SNR, $\bar{\gamma}$, of $\alpha-\eta-\mu$/gamma fading with different values of $\alpha$, $\eta$, $\mu$, and $b$ ($A = 1$).}
\end{figure} 
\begin{figure}[h]
\centering
\includegraphics[width=3.5 in, height=2.35 in]{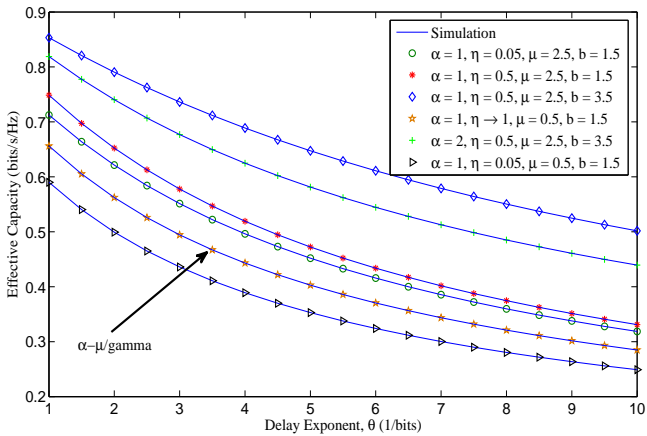} 
\centering
\caption{Simulated and numerical effective capacity using MoG distribution against delay exponent, $\theta$, of $\alpha-\eta-\mu$/gamma fading with different values of $\alpha$, $\eta$, $\mu$, and $b$ ($T = 1$ and $\bar{\gamma} = 0$ dB).}
\end{figure}
\section{Numerical results} 
Fig. 1 and Fig. 2 show the simulated and numerical effective capacity of $\alpha-\eta-\mu$/gamma fading (format 1) against the average SNR, $\bar{\gamma}$, using MG distribution and delay exponent, $\theta$, using MoG approach, respectively. The number of components for both distributions, namely, $S$ and $N$, are chosen to achieve MSE$\leq10^{-8}$. In Fig. 2, the parameters have been calculated by following the same procedure in [9]. From both figures, it can be observed that the effective capacity becomes better when $\mu$, $\eta$ or/and $b$ increase. This is because higher $\mu$, $\eta$ and $b$ mean the number of multipath clusters is large, the received power is high and the shadowing impact is low, respectively.

\section{Conclusion}
In this letter, we have used MG and MoG distributions to analyse the effective capacity over $\alpha-\eta-\mu$/gamma fading channels. These distributions can be employed to approximate with high accuracy the PDF of a wide range of distributions that are used in modelling the wireless channels. Although the MG distribution leads to simple expression, its not applicable for all fading channels. Therefore, we have utilised the MoG distribution. To this effect, unified simple closed-form mathematically tractable expressions are derived. The results have showed different scenarios that have not been yet investigated in the technical literature such as $\eta-\mu$ and $\alpha-\mu$/gamma fading channels.

\vskip3pt
\copyright
\vskip5pt

\noindent Hussien Al-Hmood (\textit{Electrical and Electronics Engineering Department, University of Thi-Qar, Thi-Qar, Iraq})\\
\noindent E-mail: hussien.al-hmood@{brunel.ac.uk, eng.utq.edu.iq}\\
H. S. Al-Raweshidy (\textit{Electronic and Computer Engineering Department, College of Engineering, Design and Physical Sciences, Brunel University London, UK})

\end{document}